\documentclass[prl,nobalancelastpage,twocolumn,superscriptaddress]{revtex4-1}

\usepackage{color,amsmath,amsfonts,graphicx,bm}
\usepackage{times}

\newcommand{\ket}[1]{\vert#1\rangle}
\newcommand{\bra}[1]{\langle#1\vert}
\newcommand{\tr}{\mathrm{tr}}
\newcommand{\mc}[1]{\mathcal{#1}}

\begin{document}

\title{An order parameter for symmetry-protected phases in one dimension}

\author{Jutho Haegeman}
\affiliation{Vienna Center for Quantum Science and Technology, Faculty of
Physics, University of Vienna, Boltzmanngasse 5, A-1090 Wien, Austria}
\author{David P\'erez-Garc\'ia}
\affiliation{Dpto.\ Analisis Matematico and IMI, 
    Universidad Complutense de Madrid, 
    E-28040 Madrid, Spain}
\author{Ignacio Cirac}
\affiliation{Max-Planck-Institut f\"ur Quantenoptik, 
    Hans-Kopfermann-Str.\ 1, 
    D-85748 Garching, Germany}
\author{Norbert Schuch}
\affiliation{Institut f\"ur Quanteninformation, RWTH Aachen,
    D-52056 Aachen, Germany}
\affiliation{Institute for Quantum Information, 
    California Institute of Technology, 
    MC 305-16, Pasadena CA 91125, U.S.A.}

\begin{abstract}
We introduce an order parameter for symmetry-protected phases
in one dimension which allows to directly identify those phases.  The
order parameter consists of string-like operators and swaps, but differs
from conventional string order operators in that it only depends on the
symmetry but not on the state.  We verify our framework through numerical
simulations for the $\mathrm{SO}(3)$ invariant spin-$1$
bilinear-biquadratic model which exhibits a dimerized and a Haldane phase,
and find that the order parameter not only works very well for the
dimerized and the Haldane phase, but it also returns a distinct signature
for gapless phases.  Finally, we discuss possible ways to measure the
order parameter in experiments with cold atoms.
\end{abstract}

\maketitle

Symmetries play an essential role almost everywhere in physics. A
prominent example is Landau's theory of phases: Different phases are
classified by whether the state of the system obeys or breaks the
symmetries of the Hamiltonian. In turn, this gives rise to local order
parameters which can be measured to determine which phase a system is in.
This picture has recently been challenged by the discovery of
topologically ordered phases: They are not associated to the breaking of
any local symmetry, and therefore there is no local order parameter which
can be used to detect topological phases.  

Topological order occurs only in two and higher dimensions, since
one-dimensional gapped spin systems exhibit only a single phase. However,
the situation in one dimension changes if we impose symmetries---they do
not only give rise to Landau-type phases which are distinguished by local
order parameters,
but also to distinct \emph{symmetry-protected phases} which cannot be
distinguished by any local order parameter (and might thus be called
\emph{topological}), yet which are protected (i.e.,
separated) by the presence of the symmetry. The most prominent example for
a non-trivial symmetry protected phase is the Haldane phase, which
contains the spin-$1$ AKLT model and likely the spin-$1$ Heisenberg model
and which is e.g.\ protected by $\mathrm{SO}(3)$
symmetry~\cite{pollmann:1d-sym-protection-prb}.
More recently, it has been realized that these phases differ by the way
in which the symmetry acts \emph{across} blocks of the system, i.e., on
the entanglement between blocks.  This can be understood in a particularly
natural way in the framework of Matrix Product States (MPS), which provide
the appropriate framework for the description of gapped one-dimensional
systems, and which allow to directly access the entanglement between
blocks~\cite{pollmann:1d-sym-protection-prb,chen:1d-phases-rg,%
schuch:mps-phases}.  In particular, it has been found that the action of
the symmetry on the entanglement between blocks, and thus the different
symmetry protected phases, are distinguished by the inequivalent
projective representations of the symmetry group, such as integer and
half-integer spin representations in the case of the rotation
group~\cite{pollmann:1d-sym-protection-prb,chen:1d-phases-rg}.

Symmetry protected phases such as the spin-$1$ AKLT chain do not exhibit
long-range order, this is, non-decaying correlations between distant
sites, which could otherwise replace local order parameters. However, they
do exhibit what is known as string
order~\cite{nijs:string-order,kennedy:string-order}: Measuring a string of
identical operators with distinct endpoints gives correlations which do
not depend on the length of the string, despite the absence of
conventional long-range order. This might suggest that string order
parameters can be used to distinguish different symmetry-protected phases.
However, the presence of string order is rather a signature of the
symmetry itself than of the phase of the system under that symmetry, and
string order parameters need to be tailored to the system under
consideration; indeed, one can easily find examples of systems in
different phases which are susceptible to the same string order parameter,
and vice versa~\cite{perez-garcia:stringorder-1d}.  Therefore, string
order parameters are not well suited 
 as order parameters for symmetry protected phases.

In this paper, we propose an order parameter which allows to distinguish
symmetry-protected phases by directly measuring the way in which the
physical symmetry acts on the entanglement between blocks.  Unlike string
order parameters, it is independent of the state under consideration and
only depends on the symmetry itself.  We demonstrate our approach for the
$\mathrm{SO}(3)$ symmetry by numerically studying  the spin--$1$
bilinear--biquadratic model (see,
e.g.,~\cite{laeuchli:bilinear-biquadratic-nematic}), where we find that
the order parameter, though defined for asymptotically large blocks,
converges very well for small lengths. We also find that while the order
parameter is designed to work for the gapped phases of the system, namely
the trivial and the Haldane phase, it in fact also exhibits a distinct
signature for the gapless critical and ferromagnetic phase of the model.
Finally, we discuss how one could in principle experimentally implement a
measurement of this order parameter, in particular with atoms in optical
lattices.

\begin{figure}[t]
\includegraphics[width=0.8\columnwidth]{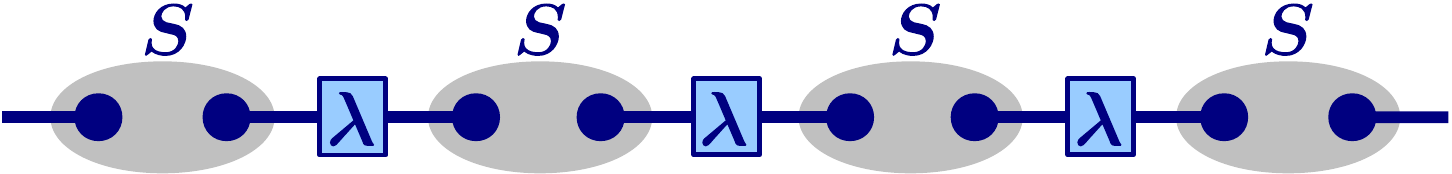}
\caption{\label{fig:rg-fp}
Large-scale structure of 1D quantum states. The renormalization fixed
point consists of virtual entangled pairs with Schmidt spectrum $\lambda$
between adjacent sites, which are mapped by an isometry $S$ onto the
physical system. If $S$ maps onto $L$ sites of the original
state, this ansatz approximates any ground state of a gapped Hamiltonian
to an accuracy exponential in $L$.}
\end{figure}

Let us start by explaining the structure of one-dimensional gapped phases
under symmetries; for simplicity, we will focus on systems with unique
ground states. As such phases differ by their long-range properties, we
first describe the structure of such states on large length scales,
and in particular their renormalization group (RG) fixed points. Due
to the absence of topological entanglement in one dimension, at the RG
fixed point of 1D systems each site independently shares entanglement
with its two adjacent sites. This is, the overall state is of the form
\begin{equation}
        \label{eq:rg-fp}
\ket{\Psi}=S^{\otimes N}\ket{\lambda}^{\otimes N}\ .
\end{equation}
Here, $\ket\lambda=\sum \lambda_i \ket{i,i}$ is a ``virtual'' entangled
state between adjacent sites with Schmidt spectrum
$\lambda=(\lambda_1,\dots,\lambda_D)$, and $S$ is an isometry mapping the
virtual entangled states into the physical system, acting jointly on the
halves of two adjacent $\ket\lambda$ states, as depicted in
Fig.~\ref{fig:rg-fp}.  In the framework of Matrix Product States (MPS),
which are obtained by replacing the isometry $S$ in Fig.~\ref{fig:rg-fp}
by an arbitrary linear map, and which form the appropriate class of states
for the description of one-dimensional gapped quantum
systems~\cite{hastings:arealaw,verstraete:faithfully,schuch:mps-entropies},
it can be proven rigorously that any ground state of a gapped 1D system
converges exponentially to the fixed point
of Eq.~(\ref{eq:rg-fp})~\cite{verstraete:renorm-MPS}; thus,
Eq.~(\ref{eq:rg-fp}) can equally well be understood as an approximation to
any gapped 1D system, where $S$ embeds the
virtual entangled pairs into a block of $L$ physical sites each, with an
accuracy exponential in $L$. (In this case, $S$ can be understood as the RG
transformation on a block of length $L$).

Now consider a quantum system with an on-site linear symmetry $u_g$,
$u_g^{\otimes N}\ket{\Psi}=\ket\Psi$, where $u_g$ is a representation of
the symmetry group $G$, $u_gu_h=u_{gh}$. Under renormalization, the
symmetry action transforms to the action $U_g$ on the renormalized sites.
(In particular, if blocking $L$ sites, $U_g=u_g^{\otimes L}$.) In the
representation of Eq.~(\ref{eq:rg-fp}) and Fig.~\ref{fig:rg-fp}, this
symmetry can be understood as an effective symmetry $\hat U_g = S^{-1} U_g
S$ acting on the virtual entangled pairs. Note that $\hat U_g$ forms again
a linear unitary representation of $G$, as $U_g$ commutes with
$SS^{\dagger}$~\cite{schuch:mps-phases}.  It can be shown that the virtual
action of the symmetry $\hat U_g$ always decomposes as $\hat
U_g=V_g\otimes \bar V_g$, where $V_g$ and $\bar V_g$ act on the left and
the right entangled state,
respectively~\cite{perez-garcia:stringorder-1d}.  Moreover,
$\Lambda:=\sum \lambda_i\ket{i}\!\bra{i}$ commutes with $\bar
V_g$~\cite{schuch:mps-phases}, so that $(\bar V_g\otimes
V_g)\ket{\lambda}=\ket{\lambda}$. 

Since $\hat U_g=V_g\otimes \bar V_g$, and $\hat U_g\hat U_h=\hat U_{gh}$,
it follows that $V_g$ forms a projective representation, 
$V_gV_h=e^{i\omega(g,h)}V_{gh}$. Here, $\omega(g,h)$ is a \emph{2-cocycle},
i.e., it satisfies 
$\omega(g,hk)+\omega(h,k)=\omega(g,h)+\omega(gh,k)\mathrm{\ mod\,}2\pi$. 
  As $V_g$
is only defined up to its phase (and up to a similarity transform), we
have a gauge degree of freedom $V_g\leftrightarrow e^{i\phi_g} V_g$ which
induces an equivalence relation
$\omega(g,h)\sim \omega(g,h)+\phi_g+\phi_h-\phi_{gh} \mathrm{\
mod\,}2\pi$ of $2$-cocycles, and thus equivalence classes of projective
representations. These equivalence classes form a group 
isomorphic to the second cohomology group $\mathrm{H}^2(G,\mathrm{U}(1))$,
and label the inequivalent projective representations of the symmetry
group $G$. For the
rotation group $\mathrm{SO(3)}$, e.g., the inequivalent projective
representations are the integer and half-integer spin representations,
respectively. It turns out that the equivalence class of the projective
representation $V_g$, which describes the action of the symmetry on the
virtual degrees of freedom, is exactly what labels different phases in the
presence of symmetries~\cite{chen:1d-phases-rg,schuch:mps-phases}.

\begin{figure}[t]
\includegraphics[width=\columnwidth]{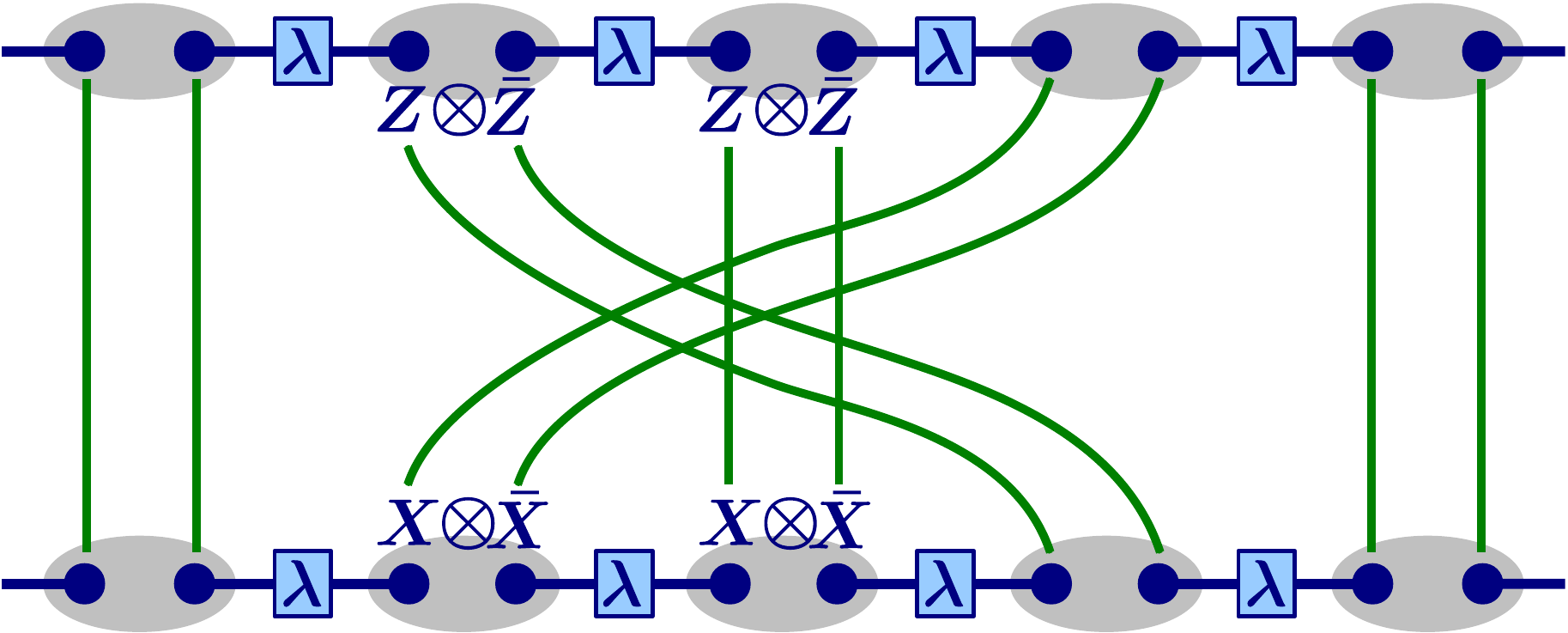}
\caption{
    \label{fig:measurement}
Illustration of the measurement Eq.~(\ref{eq:estimator}) used to determine
phases under $\mathrm{SO}(3)$ symmetry (with shorthand $Z$ for $V_z$,
etc.).  By following the loops created by the operator and the entangled
states, it can be easily checked that the diagram evaluates to $\mc
E(\Psi)$, Eq.~(\ref{eq:xzxz}); note that operators traversed downwards
have to be transposed.  } \end{figure}

In order to detect different symmetry protected phases, we therefore need
a measurement which allows us to determine the equivalence class of the
projective representation with which the symmetry acts on the entanglement
between blocks. However, the problem is that while we know $U_g$, it is
impossible to infer sufficient information about $V_g$ from it---to do so
we would have to know the transformation $S$, which would require
full tomography of the
state~\footnote{E.g., if for $\mathrm{SO(3)}$ $U_g$ contains both the
spin-$0$ and spin-$1$ representation,  $V_g\otimes \bar V_g$ could act
either as $\tfrac12\otimes \tfrac12$ or as $0\otimes 0$, depending on
$S$.}. Fortunately, we do not need detailed knowledge of
$V_g$, since we only want to know to which equivalence class of projective
representations it belongs. For this purpose, it is sufficient to compute
certain gauge invariant quantities which give access to the gauge invariant
universal signature of $\omega(g,h)$: For instance, for $\mathrm{SO}(3)$
symmetry, one such quantity is $V_zV_xV_z^\dagger
V_x^\dagger$, where $x,z\in\mathrm{SO(3)}$ denote $\pi$ rotations about
the $x$ and the $z$ axis: It is $+\openone$ for integer spin
representations and $-\openone$ for half-integer spin representations of
$\mathrm{SO}(3)$, respectively, and does not depend on the gauge. (See
Appendix A for other groups.) Thus, if
we were able to measure such an invariant for a given state, we would
be able to determine the symmetry protected phase the system is in.
Different from determining $V_g$ itself, this invariant can be determined
without any information about $S$, by measuring a suitable operator
such as
\begin{equation}
        \label{eq:estimator}
\mathcal E(\Psi):=\bra\Psi (U_z\otimes U_z\otimes\openone) \mathbb F_{13}
(U_x\otimes U_x\otimes\openone) \ket{\Psi} \ ,
\end{equation}
where $\mathbb F_{13}$ swaps the first and the third site. 
[Equivalently, $\mc E(\Psi)=\bra\Psi \mathbb F_{13}
U_x U_y U_z\ket\Psi$.] If one
expresses this measurement diagrammatically using that $\hat
U_g=V_g\otimes \bar V_g$, cf.~Fig.~\ref{fig:measurement},
it is straightforward to check that 
\begin{equation}
        \label{eq:xzxz}
\mathcal E(\Psi) = \tr[V_zV_xV_z^\dagger V_x^\dagger \Lambda^4]\,
    \tr[\Lambda^4]\ ,
\end{equation}
and its sign thus allows to determine the phase of $\ket\Psi$.
 (Recall that $\Lambda$ commutes with $V_g$.) Note that by omitting
the $U$'s in Eq.~(\ref{eq:estimator}), one can measure $\mathcal
N(\Psi):=\tr[\Lambda^4]^2$, and that the ratio $\hat{\mc E}(\Psi):=
\mc E(\Psi)/\mc N(\Psi)=\pm 1$ yields a normalized quantity which
distinguishes the trivial from the Haldane phase. (See Appendix B for how
to measure general gauge invariant quantities.)

The preceding discussion was concerned with
renormalization fixed points, i.e., states of the form
Eq.~(\ref{eq:rg-fp}).  However, we can evaluate the same quantity for 
an arbitrary quantum state, by replacing $U_x$ and $U_z$ by strings of
local symmetry operators $u_x$ and $u_z$, i.e., $U_g=u_g^{\otimes L}$. For
ground states of gapped Hamiltonians, we expect it to converge
exponentially fast to the value at the renormalization fixed
point; this can be proven rigorously in the framework of Matrix Product
States, cf.~Appendix C.

In order to test the applicability of the order parameter, we have
performed numerical simulations for the spin-$1$ bilinear-biquadratic
Heisenberg chain (cf.,
e.g.,~\cite{laeuchli:bilinear-biquadratic-nematic} and references therein)
\begin{equation}
    \label{eq:bilin-biquad}
H(\theta) = \cos\theta \sum_i {\bm S}_i \cdot {\bm S}_{i+1}
	+\sin\theta \sum_i (\bm S_i \cdot \bm S_{i+1})^2\,.
\end{equation}
This model is $\mathrm{SO}(3)$ invariant and exhibits both 
possible
gapped phases under rotational symmetry: A dimerized phase for
$-3\pi/4 < \theta < -\pi/4$ (with integer spin representations $V_g$ and
thus topologically trivial), and a Haldane phase for
$-\pi/4 < \theta < \pi/4$ (with half-integer representations
$V_g$ and thus topologically non-trivial).

\begin{figure}[t]
        \includegraphics[width=0.95\columnwidth]{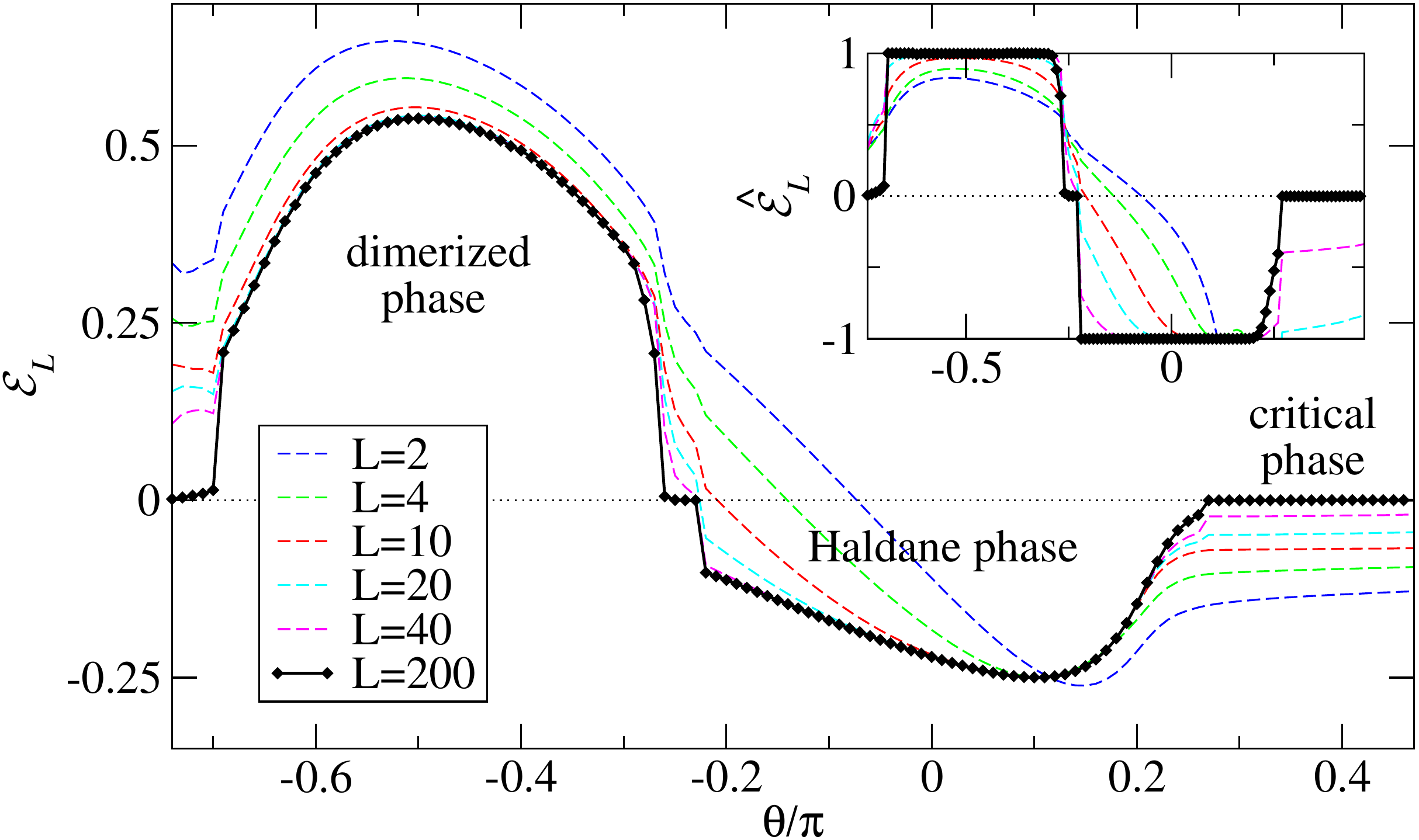}
\caption{
\label{fig:estim-data}
Order parameter $\mc E_L$ for the bilinear-biquadratic spin-$1$ model,
Eq.~\eqref{eq:bilin-biquad}, as a function of $\theta$, for different
estimator lengths $L$. The inset shows the normalized order parameter
$\hat{\mc E}_L$. Note that $-1/4\le\mc E\le1$ from the minimal
dimension of the corresponding representations. $\mc E=-1/4$ is attained
at the ALKT point ($\theta/\pi\approx 0.1024$).
}
\end{figure}

The simulations have been carried out using infinite Matrix Product States
(iMPS)~\cite{vidal:iTEBD} with sites blocked in pairs, using the time-dependent
variational principle~\cite{haegeman:varprinciple-lattices}.  The results
for the order parameter ${\mc E_L}$ as a function of $\theta/\pi$ are
plotted in Fig.~\ref{fig:estim-data} for different length $L$ 
($L$ refers to the length of a single block in
Fig.~\ref{fig:measurement}), and for $D=32$; the inset shows the
normalized $\hat{\mc E_L}$.  We find that $\mc E_L$ converges quickly,
with its sign correctly distinguishing the dimerized phase ($+1$) from the
Haldane phase ($-1$).  Deviations from this behavior can be observed at
the phase transitions around $\theta=\pm\pi/4$,
 as well as in the dimerized phase close to
$\theta=-3\pi/4$, a regime in which the possible existence of a
``spin-nematic phase'' is under ongoing debate~(see, e.g.,
\cite{laeuchli:bilinear-biquadratic-nematic}).  Simulations with different
$D$ give similar results, except that the width of the transition regions
between the phases decreases with increasing $D$.

\begin{figure}
    \includegraphics[width=\columnwidth]{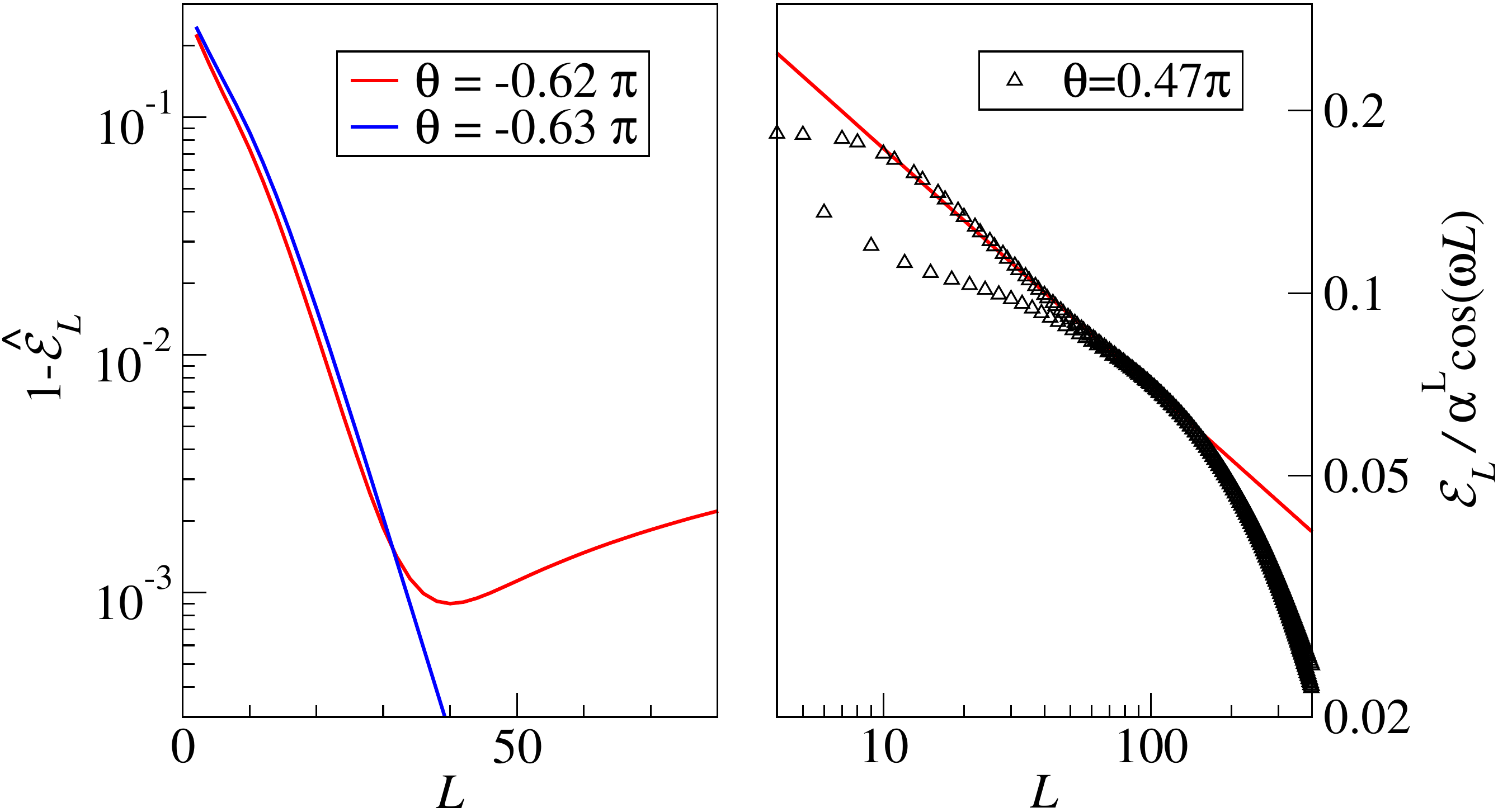}
    \caption{\label{fig:convergence}
Convergence analysis. \textbf{a)} Normalized order parameter $\hat{\mc
E}_L$ for $\theta/\pi=-0.62$ and $-0.63$. Initially, both converge
to $1$ exponentially with similar exponent, but for $\theta/\pi=-0.62$,
the convergence is not perfect as $D$ is incommensurate with the
representation of the symmetry.  \textbf{b)} Algebraic decay in the
critical phase.  After correcting for the exponential decay from the iMPS
approximation and oscillations, we find that the estimator goes to zero
algebraically.  For the fit, we have only used $L$ with
$\mathrm{mod}(L,3)\ne 0$.  }
\end{figure}

A closer analysis of the data reveals that inside both phases, $\mc E_L$
converges exponentially  with a length scale
essentially equal to the correlation length (deviation below $2\%$).
However, for many values of $\theta$ this behavior is only seen on
intermediate length scales (typically up to $L\approx 30\dots 40$), while
at larger scales, $\mc E_L$ tends to zero. This can be understood as
follows: Finding the optimal MPS approximation with a given $D$
corresponds to keeping the $D$ largest values in the Schmidt spectrum
$\lambda$ of the state.  As $[\Lambda,V_g]=0$, the degeneracies
in the Schmidt spectrum correspond to the irreps of $\mathrm{SO}(3)$
which appear in $V_g$.  If the truncation does not respect these 
degeneracies
(this depends on the ordering of the irreps in the Schmidt spectrum and
thus on the point $\theta$ in the phase), the resulting iMPS will not any
more be exactly $\mathrm{SO}(3)$ invariant, which causes $\mc E_L$ to
converge to zero.
For the data reported in Fig.~\ref{fig:estim-data} with $D=32$, we find
perfect convergence for $\theta/\pi=-0.69\,\dots{-0.63}$ and
$\theta/\pi=-0.22\dots{}{-0.10}$. In Fig.~\ref{fig:convergence}a, we
compare the two cases for two adjacent points in the dimerized phase.

Beyond the dimerized and the Haldane phase, the bilinear-biquadratic model
also exhibits two gapless phases. Firstly, a ferromagnetic phase for
$\pi/2\!\le\theta\le 5\pi/4$ with product ground states
$\ket{\phi}^{\otimes N}$; there, $\mc E_L=
(\bra\phi u_x\ket\phi\bra\phi
u_y\ket\phi\bra\phi u_z\ket\phi)^L$ converges to
zero exponentially in $L$.  Secondly, there is a critical phase in the
regime $\pi/4 \le \theta <
\pi/2$~\cite{laeuchli:bilinear-biquadratic-nematic} which we have included
in our simulations, cf.~Fig.~\ref{fig:estim-data}.  The results for this
region should be taken with care, as MPS perform
considerably worse in describing the ground states of critical systems,
and in particular cannot exactly reproduce algebraically decaying
correlations.  Analyzing the behavior of the order parameter in the critical
regime, we find a dominating exponential decay due to the fact that $D$ is
incommensurate with the symmetry, which is superimposed with an oscillation
due to the $2\pi/3$ periodicity of the critical
phase~\cite{laeuchli:bilinear-biquadratic-nematic}, $\alpha^L \cos(\omega
L)$. The parameters $\alpha$ and $\omega$ can be extracted from the
MPS transfer operator, and we find that after correcting for these
effects, $\hat{\mc E}_L/(\alpha^L \cos(\omega L))$ (plotted in
Fig.~\ref{fig:convergence}b for $\theta=0.47$) exhibits an algebraic decay
to zero, with an exponent which varies between $0.41$ and $0.44$ in the
critical phase. The results on the exponent should be taken with
particularly great care, as they depend on the choice of data points
fitted, and the algebraic decay observed is in fact a sum of 
exponentials; yet, we believe that our results provide
substantial evidence for an algebraically vanishing order parameter in the
critical phase.

So far, we have shown that the order parameter ${\cal E}(\Psi)$ can be
used as a theoretical tool to determine the phases of 1D spin
chains.  Now we will show that, at least in principle, it can also be
experimentally determined by performing few measurements (without the need
of carrying out a full tomography). The main idea is to use an ancillary
particle which controls whether the unitary and swapping operations
appearing in Eq.\ (\ref{eq:estimator}) are applied or not, and then
perform a measurement on that particle. Let us assume that the ancilla is
a qubit, initially prepared in the state $(\ket{0}+\ket{1})/\sqrt{2}$.
Then, the ancilla interacts successively with particles in region 1 and 2,
1 and 3, and then again 1 and 2, such that if it is in state $\ket{1}$,
the unitary operator appearing in Eq.\ (\ref{eq:estimator}) is applied,
and otherwise just the identity operator.  At the end of the process, one
measures the Pauli operator $\sigma_x$ on the ancilla, whose expectation
value conincides with ${\cal E}(\Psi)$.

The techniques required to carry out the above procedure are very
sophisticated, and we do not expect that they can be performed in most of
the experimentally relevant situations (beside atomic physics experiments,
where individual addressing and full control over the atomic interactions
may be gained in the near future
\cite{greiner:atomaddressing,bloch:atomaddressing,bloch:stringorder}).  In
any case, here we give an alternative method to detect the phase which may
be slightly simpler to implement in the particular setup of atoms in
optical lattices. The main idea is to use two copies of the spin chain, as
it is usual in that setup (for instance, with the help of superlattices
\cite{lewenstein:reviewlattices}). Then, one would like to determine
\begin{align}
{\cal E}^{(2)}(\Psi) := &
    \bra{\Psi}_A\bra{\Psi}_B 
    (\mathbb F_{1A,1B}\otimes \mathbb F_{3A,3B})\,\times
    \label{eq:two_copy_estimator}
\\
&\quad
    (U_z^{1,A}\otimes U_z^{1,B}\otimes
    U_x^{2,A}\otimes U_x^{2,B}) \ket{\Psi}_A\ket{\Psi}_B\ .
    \nonumber
\end{align}
Here, $A$ and $B$ refer to the first and second copy of the chain, and 1,
2, and 3 to three neighboring regions each containing a sufficiently large
number of spins. One can easily convince oneself that ${\cal E}^{(2)}$
contains the same information as ${\cal E}$. 
 The advantage of this measurement is that the
swap only occurs between particles in two chains which are adjacent to
each other.  In practice, the ancilla could consist of a different atomic
species (see \cite{Aguado:ancillaatoms}), so that it can be transported
independently of the spin chains. As explained in that reference, one
could use this ancilla to apply the conditional unitary operators
sequentially to the spin chains. Additionally, the swapping operator can
be generated by letting the ancilla control the pairwise interaction among
the neighboring spins of the first and second chain in regions 1 and 3,
corresponding to a Hamiltonian $H=\sum (h_i+h_i^2)$, where
$h_i={\bm S}_{i,A} \cdot {\bm S}_{i,B}$, for a time $t=\pi/2$. We would
like to emphasize that this procedure may be very difficult in practice,
but it still shows that in principle one can measure the order parameter.

To conclude, in this paper we have introduced an order parameter for
symmetry protected phases in one dimension.  We have illustrated our
construction for $\mathrm{SO(3)}$ symmetry, where we have verified our
predicitions numerically for the spin-$1$ bilinear-biquadratic model.  We
found that the order parameter allows to faithfully determine which gapped
phase the system is in; moreover, we found that (somewhat surprisingly)
it also returns a distinct signature for the gapless phases of the
model.

Similar order parameters can be constructed for symmetry protected gapped
phases with partial symmetry breaking~\cite{schuch:mps-phases},
 by first using conventional local order parameters to
detect which symmetries are broken, and subsequently measuring order
parameters such as the one presented to detect the topological phase
protected by the remaining unbroken symmetries.  An interesting open
question is whether our method to
identify equivalence classes of $2$-cocycles, corresponding to elements in
$\mathrm{H}^2(G,\mathrm{U}(1))$, can be modified to distinguish symmetry
protected phases in two dimensions, which are labelled by equivalence
classes of $3$-cocycles and correspondingly the third cohomology group
$\mathrm{H}^3(G,\mathrm{U}(1))$. Finally, as the endpoints of string
operators can be interpreted as quasi-particles, it would be interesting to
understand whether our order parameter can be effectively understood as
extracting information about the quasi-particle braiding properties.

\emph{Acknowledgments.---}We thank J.\ de Boer and F.\ Verstraete for
helpful comments. Parts of this work were carried out at the Perimeter
Institute in Waterloo, Canada, and the Centro de Ciencias Pedro Pascual in
Benasque, Spain.  This work has been supported by the Alexander von
Humboldt Foundation, the Caltech Institute for Quantum Information and
Matter (an NSF Physics Frontiers Center with support of the Gordon and
Betty Moore Foundation), the NSF Grant No.~\mbox{PHY-0803371}, the EU
grants QUEVADIS and QUERG, the Spanish projects QUITEMAD and
MTM2011-26912, the FWF SFB grants \mbox{FoQuS} and ViCoM, and the
DFG-Forschergruppe 635.

\appendix

\section{Appendix A: Distinguishing inequivalent projective representations}

As we show in Appendix B, we can measure any gauge-invariant quantity of
the form 
\begin{equation*}
\tr[\tilde V_g\tilde V_h\dots]\, \times\,
\tr[\tilde V_k\tilde V_\ell\dots]\,\times\cdots\ ,
\end{equation*}
where $\tilde V_g$ is either $V_g\Lambda$ or $V_g^\dagger\Lambda$, and for
each $g$, $V_g$ and $V_g^\dagger$ appear an equal number of times. Since
$[V_g,\Lambda]=0$, we can (and will) omit the $\Lambda$'s in the
following.  As we are interested in gauge-invariant 
 observables, the product of the group elements
in each trace must be equal to the identity, and it follows that it is
sufficient to consider a single trace. (Note that this is all we can
measure, since $V_g$ is only defined up to a phase and similarity
transform, $V_g\leftrightarrow e^{i\phi_g}X_gV_gX_g^{-1}$.)

Let us now discuss some cases beyond $\mathrm{SO}(3)$ in which
we can identify the equivalence class of the projective representation by
such a measurement.  Most importantly, this holds for any special
orthogonal group $\mathrm{SO}(N)$, $N\ge3$, all of which have a two-fold
covering by the corresponding spin group $\mathrm{Spin}(N)$, and 
correspondingly $\mathrm{H}^2(\mathrm{SO}(N),\mathrm{U}(1))=\mathbb Z_2$.
As for $\mathrm{SO}(3)$, $V_xV_zV_x^\dagger V_z^\dagger=\pm\openone$, with 
$x$ and $z$ $\pi$-rotations about two orthogonal axes, will allow to
identify the equivalence class of $V_g$. (This can e.g.\ be understood by
considering projective representations of the subgroup $\mathbb Z_2\times
\mathbb Z_2$ generated by $x$ and $z$.)  Furthermore, the same result
holds for any subgroup of $\mathrm{SO}(N)$ containing the group elements
which appear in the measurement, such as $x$ and $z$. We expect similar
measurements to exist for other finite groups.

\section{\label{sec:app-b}
Appendix B: Measuring general gauge-invariant quantities}

\begin{figure}[h]
\includegraphics[width=\columnwidth]{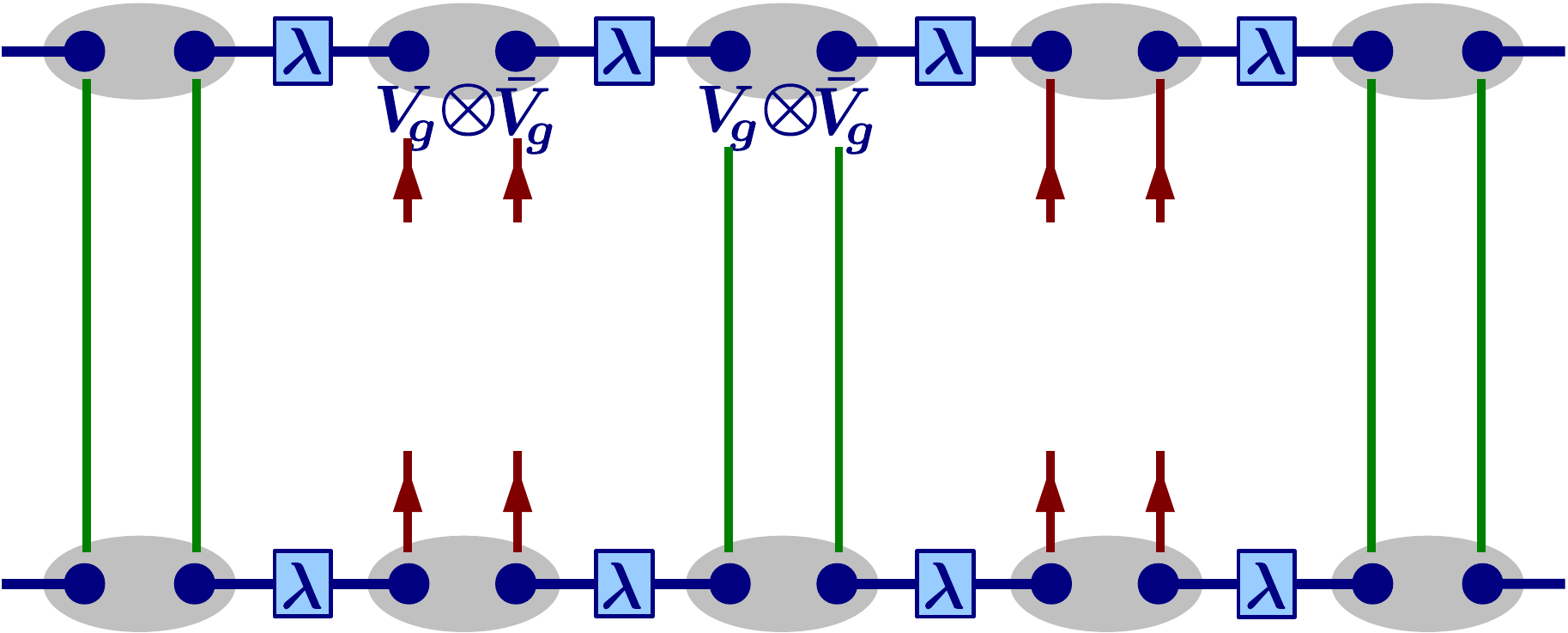}
\caption{\label{fig:app-b-gadget}
Elementary gadget used to construct arbitrary gauge invariant
measurements.  It has two input sites and two output sites (each
consisting of a left and a right subsystem), which can be
arbitrarily wired together by swapping sites.
}
\end{figure}

In this appendix, we show how to measure an arbitary gauge invariant
quantity of the form
\begin{equation}
\label{eq:app-b:ginv-trace}
\tr[\tilde V_g\tilde V_h\dots]\, \times\,
\tr[\tilde V_k\tilde V_\ell\dots]\,\times\cdots\ ,
\end{equation}
where $\tilde V_g$ is either $V_g\Lambda$ or $V_g^\dagger\Lambda$, 
the product in each trace is proportional to the identity (allowing us to
disregard $\Lambda$'s) and for each $g$, $V_g$ and $V_g^\dagger$ appear an
equal number of times, ensuring gauge invariance.  The measurement is
implemented by concatenating the elementary ``gadgets'' of the form of
Fig.~\ref{fig:app-b-gadget}, which we think of as objects with two
``inputs'' (the top row) and two ``outputs'' (the bottom row); the outputs
are related to the inputs by multiplication with
$(V_g\Lambda)\otimes\Lambda$ and $(V_g^\dagger\Lambda)\otimes\Lambda$,
respectively.  It is now easy to see that arranging the gadgets for all
required $V_g$ one after another, and appropriately connecting inputs and
outputs (corresponding to a permutation of the input/output sites), the
observable Eq.~(\ref{eq:app-b:ginv-trace}) can be measured.

\section{Appendix C: Convergence to RG fixed point}

Let us show that for MPS, $\mc E_L$ converges exponentially to the value
at the RG fixed point. To this end, let $A^i$ be the MPS matrices, 
implicitly defined via $S=\sum
(A^i)_{\alpha\beta}\ket{i}\bra{\alpha,\beta}$, where the gauge 
$A^i\leftrightarrow XA^iX^{-1}$ is chosen
according to Ref.~\cite{perez-garcia:mps-reps}.
Define the transfer
operators $\mathbb E=\sum \bar A^i\otimes A^i$ and 
\[
\mathbb E_g = \sum (u_g)_{ij} \bar A^i\otimes A^j = 
(\openone\otimes V_g) \mathbb E (\openone\otimes V_g)^\dagger\ .
\]
With $\mathbb T = \mathbb E^L$ and
$\mathbb T_g = \mathbb E_g^L$, we have that 
\[
\mathcal E_L(\psi) = \sum \mathbb L^{ii'} (\mathbb T_x)_{ik'}^{j\ell'}
(\mathbb T_{xz})_{jj'}^{kk'} (\mathbb T_z)_{ki'}^{\ell j'} \mathbb
R_{\ell\ell'} 
\]
where lower (upper) indices denote the left (right) indices of the
transfer operators, and $\mathbb L=\Lambda$, $\mathbb R=\openone$ are the
unique left and right fixed points of $\mathbb
T$~\cite{perez-garcia:mps-reps}. The normalization
$\mathcal N_L(\psi)$ is obtained by replacing all $\mathbb T_g$ by $\mathbb
T$. As $\mathbb T_g=(\openone \otimes V_g)\mathbb T (\openone\otimes
V_g)^\dagger$, and $\mathbb T$ converges to its fixed point
$\ket{\openone}\bra{\Lambda}$
exponentially in $L$, exponential convergence of $\mc E_L$ and $\mc N_L$
follows.  Note that the length scale of convergence is given by the gap of
$\mathbb T$ and is thus equal to the correlation length.

\end{document}